\documentclass[%
 reprint,
 amsmath,amssymb,
 aps,
pra,
]{revtex4-2}

\usepackage{graphicx}% Include figure files
\usepackage{dcolumn}% Align table columns on decimal point
\usepackage{bm}
\usepackage{amsmath}
\usepackage{nicematrix}
\usepackage{physics}
\usepackage{comment}
\usepackage{subfigure}
\usepackage{hyperref}
\usepackage{tikz}
\usetikzlibrary{decorations.pathmorphing}
\usepackage{graphicx}
\usepackage{gnuplot-lua-tikz}
\usepackage{float}% bold math
\begin{document}

\preprint{APS/123-QED}

\title{Wichmann-Kroll vacuum polarization correction to lithium-like systems \\ in a Gaussian basis set}% Force line breaks with \\
%\thanks{hhayat@student.unimelb.edu.au}%

\author{Haisum Hayat}
\email[Contact: ]{hhayat@student.unimelb.edu.au}
\author{Harry M. Quiney}%

\affiliation{%
 School of Physics, University of Melbourne
}%


%\affiliation{
%\affiliation{
%\affiliation{%


\date{\today}% It is always \today, today,

\begin{abstract}
Recent developments have seen the application of finite Gaussian basis sets to the $\alpha(Z\alpha)^{n\geq3}$ vacuum polarization. The energy shift for $s$ and $p$ electron states have been tabulated and their convergence investigated. In this work, we extend this problem to the multi-electron case. Hartee-Fock potentials obtained self-consistently are used to treat the vacuum polarization for lithium-like systems and are found to be in good agreement with comparable results in the literature. The results presented in this work demonstrate the use of Gaussian basis sets for atomic potentials whose Green's functions expressions cannot be simply obtained via analytic or numerical methods.
%\begin{description}
%\end{description}
\end{abstract}

\maketitle

%\tableofcontents

\section{\label{sec:level1}INTRODUCTION}
The most fundamental description of the interactions within atoms and molecules lies within the theoretical framework of quantum electrodynamics (QED), which governs the relativistic and quantum behavior of charged particles and photons. Motivated by the discovery that the 2$s_{1/2}$ and 2$p_{1/2}$ states in hydrogen are non-degenerate \cite{Lamb}, as predicted by the Dirac equation, the theory of QED and tests of it have been at the forefront of fundamental physics since the 1940s. In weak-field experiments, precision tests of QED agree with theoretical predictions almost exactly \cite{codata,anomalous}; however, both theory and experiment of interactions in the presence of strong electromagnetic fields is limited. 

Some of the strongest electromagnetic fields known to us are produced by atomic nuclei; specifically, the nuclei of highly charged ions (HCI). QED calculations of these systems require non-perturbative methods that incorporate all orders of the coupling constant $Z\alpha$ \cite{shab2002}, where $Z$ is the nuclear charge and $\alpha$ is the fine-structure constant. Consequently, relativistic atomic structure calculations of heavy systems rarely treat QED corrections from a first principles approach, relying instead on approximations such as atomic psuedopotentials \cite{hangele_coupled-cluster_2014, LCG}. These often require fits of experimental data which, particularly in the case of HCIs, can be difficult to obtain. 

In this work, we investigate the vacuum polarization correction, which, along with self energy, is the lowest-order QED correction to an atomic electron. Calculations of this effect began in 1935 with Uehling \cite{uehling} and Serber \cite{serber}, who determined the first-order $(Z\alpha)$ contribution to the vacuum polarization long before standard renormalization techniques had been invented. 
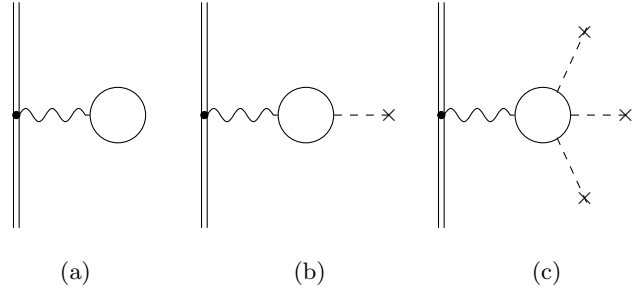
\begin{figure}[ht]
\begin{tikzpicture}[scale=0.75]%,xscale=1.4]
% full VP
%\draw (-6,-2) -- (-6,2);
%\draw (-6.1,-2) -- (-6.1,2);
%\draw (-4.25,0) circle (14pt);
%\draw (-4.25,0) circle (11pt);
%\filldraw[black] (-6.05,0) circle (1.8pt);
%\draw[decorate,decoration={coil,aspect=0}] (-6,0)  -- (-4.75,0);

%\node at (-3.1,0) {\Large =};
%\node at (0.99,0) {\Large +};
%\node at (6.05,0) {\Large +};
%\node at (11,0) {\Large +};
%\node at (12.0,0) {\Large ...};

\node at (-1.0,-2.8) {(a)};
\node at (3.15,-2.8) {(b)};
\node at (7.35,-2.8) {(c)};

% zero potential
\draw (-2,-2) -- (-2,2);
\draw (-2.1,-2) -- (-2.1,2);
\draw (-0.25,0) circle (14pt);
\filldraw[black] (-2.05,0) circle (1.8pt);
\draw[decorate,decoration={coil,aspect=0}] (-2,0)  -- (-0.75,0);

\hspace{-0.5cm}
%one potential
\draw (1.9,-2) -- (1.9,2);
\draw (2,-2) -- (2,2);
\draw (3.75,0) circle (14pt);
\filldraw[black] (1.95,0) circle (1.8pt);
\node at (5.22,0) {$\times$};
\draw[decorate,decoration={coil,aspect=0}] (2,0)  -- (3.25,0);
\draw[dashed] (4.25,0)  -- (5.25,0);

\hspace{-0.6cm}
% wichmann-kroll potential 
\draw (6.9,-2) -- (6.9,2);
\draw (7,-2) -- (7,2);
\draw (8.75,0) circle (14pt);
\filldraw[black] (6.95,0) circle (1.8pt);
\draw[decorate,decoration={coil,aspect=0}] (7,0)  -- (8.25,0);
\draw[dashed] (9.25,0)  -- (10.25,0);
\draw[dashed] (9,0.4)  -- (9.55,1.6);
\draw[dashed] (9,-0.4)  -- (9.55,-1.6);
\node at (9.49,-1.48) {$\times$};
\node at (9.49,1.48) {$\times$};
\node at (10.22,0) {$\times$};
\end{tikzpicture}
    \caption{Leading order contributions to the total unrenormalized vacuum polarization. Double lines indicate electron propagation in an external field. The individual terms are the (a) zero-potential, (b) one-potential and (c) three-potential.}
\label{fey} 
\end{figure}

Higher-order effects were first studied by Wichmann and Kroll \cite{wichmann1956} who obtained an expression for the $(Z\alpha)^3$ contribution in terms of the vacuum polarization charge density. They were able to isolate the first- and third-order contribution, showing also that the regularized version of the first-order correction is associated with the Uehling potential. The Feynman diagrams associated with this expansion are displayed in Fig.\ \ref{fey}, where the double lines indicate wavefunctions in the presence of an external Coulomb field. It should be noted that according to Furry's theorem \cite{furry} Feynman diagrams with an even number of vertices vanish; consequently, all even orders of the vacuum polarization (including the zeroth) vanish accordingly.

Due to their complexity, higher-order terms of $(Z\alpha)^{n\geq5}$ have never been derived directly. Fortunately, one can obtain all orders of $(Z\alpha)^{n\geq3}$ through subtraction of the divergent first-order Uehling term from the total unrenormalised vacuum polarization. This scheme first appeared in the work of Rinker and Wilts \cite{rinker_wilts} and Gyulassy \cite{GYULASSY1975497} who, based on the formalism of Wichmann and Kroll, performed a partial wave expansion in angular momentum $\kappa$ of the vacuum polarization charge density and calculated the $(Z\alpha)^{n\geq3}$ correction for $\abs{\kappa}=1$. The subtraction of the first-order correction also removes the physical Uehling contribution; however, this can be added back after the fact as an exact expression is known \cite{uehling}.
%\begin{align}
%    U_{\text{Ueh}}(\mathbf{x}) = -\frac{2\alpha(Z\alpha)}{3\pi}c\int \dd^3y \frac{\rho^n(\mathbf{x}-\mathbf{y})}{|\mathbf{y}|}
%        K_1\left(\frac{2|\mathbf{y}|}{\lambdabar}\right),
%        \label{uehling}
%\end{align}
%where
%\begin{align}
%    K_1(x) = \int_{1}^{\infty} \dd \zeta e^{-x\zeta}\left(\frac{1}{\zeta^2} + \frac{1}{2\zeta^4}\right) \sqrt{\zeta^2-1}.
%\end{align}

Following a similar approach to Rinker and Wilts, Soff and Mohr \cite{Morh_strong_1998} produced a set of accurate calculations of the Wichmann-Kroll correction for a range of one-electron systems, including also finite nuclear size effects. Their results still provide a useful benchmark for comparison to this day. Later, Persson \textit{et al} \cite{Persson_1993} improved on their method, providing a more numerically stable approach to the Wichmann-Kroll correction that can be applied to non-Coulomb potentials.

Vacuum-polarization screening corrections are most commonly performed using the numerical Green's function, which has been explored for helium-like \cite{He_like1,He_like2,He_like3,He_like4} and lithium-like \cite{Li_like2,VP_Lilike} systems. We note that basis set methods are yet to be investigated for these few-electron systems.
\section{\label{sec:level2}FORMALISM}

\subsection{Radial Dirac Equation}
For atomic systems one may always assume a spherically symmetric scaler potential that allows a Dirac spinor to be separated as
\begin{align}
    \psi(\mathbf{x}) = \frac{1}{r}
    \begin{bNiceMatrix}[cell-space-limits = 3pt]
        P_{n\kappa}(r) \chi_{\kappa,m}(\theta,\phi) \\
        iQ_{n\kappa}(r) \chi_{\kappa,-m}(\theta,\phi)
    \end{bNiceMatrix},
\end{align}
where $P_{n\kappa}(r)$ and $Q_{n\kappa}(r)$ are the large- and small-component radial functions respectively and $\chi_{\pm\kappa,m}(\theta,\phi)$ are the spin angular functions. In general, the angular part of the Dirac equation can be solved analytically \cite{big_grant_book,qed_book}, reducing the problem to the radial Dirac Hamiltonian
\begin{equation}
    h = 
    \begin{bNiceMatrix}[cell-space-limits = 3pt]
        c^2-V(r) & c\left(-\frac{\dd}{\dd r}+\frac{\kappa}{r}\right) \\
        c\left(\frac{\dd}{\dd r}+\frac{\kappa}{r}\right) & -c^2+V(r)
    \end{bNiceMatrix},
    \label{eq:radham}
\end{equation}
whose solutions are the radial Dirac spinors, given by
\begin{align}
    \phi_{n\kappa}(r) = \begin{bNiceMatrix}[cell-space-limits = 3pt]
        P_{n\kappa}(r) \\
        iQ_{n\kappa}(r)
    \end{bNiceMatrix}.
\end{align}
An analytic solution to Eq.\ (\ref{eq:radham}) exists only if the potential $V$ corresponds to a point-nucleus, whose charge is given by the Dirac delta distribution \cite{dirac_equation_solutions}. In any other case, numerical approaches are required.

\subsection{Dirac-Hartree-Fock}
For many-electron systems, a convenient starting point based on the independent particle model sees the wavefunction expressed as a single Slater determinant \cite{szabo_book}
\begin{align}
    \Phi = \frac{1}{\sqrt{N!}} 
    \begin{vNiceMatrix}[cell-space-limits = 3pt]
    \psi_{1}(\mathbf{x}_1) & \psi_{2}(\mathbf{x}_1) & \Cdots & \psi_{N}(\mathbf{x}_1) \\ 
    \psi_{1}(\mathbf{x}_2) & \psi_{2}(\mathbf{x}_2) & \Cdots & \psi_{N}(\mathbf{x}_2) \\
    \Vdots                 &  \Vdots                &  \Ddots    & \Vdots               \\
     \psi_{1}(\mathbf{x}_N) & \psi_{2}(\mathbf{x}_N) & \Cdots & \psi_{N}(\mathbf{x}_N)
    \end{vNiceMatrix},
    \label{Slater}
\end{align}
where $\psi(\mathbf{x})$ represents a one-electron solution to the Dirac equation and $N$ is a normalization factor. In the Dirac-Hartree-Fock (DHF) method, the energy expectation value of the Slater determinant \cite{MBPT_book}
\begin{align}
    \bra{\Phi}H_{\text{DC}}\ket{\Phi},
\end{align}
is minimized with respect to variations in $\psi(\mathbf{x})$. Here, $H_\text{DC}$ is the Dirac-Coulomb Hamiltonian \cite{DC_HAM}, which includes the electron-electron interaction through the Coulomb operator:
\begin{align}
    H_\text{DC} = \sum_i^{N_e} h_i + \sum_{i<j}^{N_e} \frac{1}{r_{ij}},
\end{align}
where $h_i$ is the familiar one-electron Dirac Hamiltonian \cite{D_HAM} and $1/r_{ij}$ is the Coulomb repulsion between electrons $i$ and $j$. The energy minimization is achieved by solving the set of coupled equations for the Fock operator
\begin{align}
    f_i = h_i + \sum_j (J_j - K_j),
\end{align}
in which the electron-electron interaction is expressed through the direct operator
\begin{align}
    J_i(\mathbf{x_1})\phi(\mathbf{x_1}) = \bra{\psi_i(\mathbf{x_2})}\frac{1}{r_{12}}\ket{\psi_i(\mathbf{x_2})}\phi(\mathbf{x_1}),
    \label{direct}
\end{align}
and the exchange operator
\begin{align}
    K_i(\mathbf{x_1})\phi(\mathbf{x_1}) = \bra{\psi_i(\mathbf{x_2})}\frac{1}{r_{12}}\ket{\phi(\mathbf{x_2})}\psi_i(\mathbf{x_1}),
    \label{exchange}
\end{align}
noting that the integration is over electron 2 only. It is clear from Eqs.\ (\ref{direct}) and (\ref{exchange}) that the eigenvalue problem
\begin{align}
    f_i\psi_i = \epsilon_i\psi_i,
    \label{fock_scf}
\end{align}
for the set of one-electron eigenfunctions depends on the eigenfunctions themselves. To obtain meaningful solutions, we iteratively solve Eq.\ (\ref{fock_scf}) until the one-electron orbitals used in the definitions of $f_i$ are consistent with the solutions $\psi_i$. The result of this self-consistent procedure describes the motion of each electron in the time-averaged field of the other electrons. Finally, we note that the Fock operator is a one-electron operator whose eigenvalues describe the energy of an electron in the field of a nucleus, screened by the other electrons in the system.

\subsection{Vacuum Polarization}
In this work, we follow the method of Persson \textit{et al} for calculations of the vacuum polarization energy correction. The complete details of the derivation can be found in Ref.\ \cite{Persson_1993}. Starting from the Feynman diagram of the total vacuum polarization, one can expand the photon propagator in partial waves
\begin{align}
    U_{\text{VP}}(\mathbf{x}_1) = &-\frac{2i}{\pi}\int d^3 \mathbf{x}_2 \int_{0}^{\infty}\dd k\int_{-\infty}^{\infty}\frac{\dd z}{2\pi}\notag\\ 
                               &\times\sum_{l=0}^{\infty} (2l+1)j_l(kr_1)\mathbf{C}^l(1)\alpha^{\mu}(1) \notag\\
                               &\times\sum_t \frac{\psi_t^{\dagger}(\mathbf{x}_2) j_l(kr_2)\mathbf{C}^l(2)\alpha_{\mu}(2)\psi_t(\mathbf{x}_2)}
                                {z-E_t(1-i\eta)},
\end{align}
where $\mathbf{C}^l(i)$ denotes a spherical tensor of order $l$ and coordinate $i$ and $j_0(r)$ is a spherical Bessel function. It can be shown that non-vanishing contributions exist only when $l$ and $\mu$ are zero. Performing the $z$ integration  gives an expression for the total unrenormalized vacuum polarization in the form of a partial wave sum
\begin{align}
    U_{\text{VP}}(r) = - &\frac{1}{\pi} \int_{0}^{\infty} \dd k~j_0(kr) \sum_{\kappa}(2j_{\kappa}+1) \notag \\
                    &\times\sum_n \text{sgn}(E_{n\kappa,Z})\bra{n\kappa,Z}j_0(kr_1)\ket{n\kappa,Z},
    \label{unrenormalized_vapol}
\end{align}
where $\ket{n\kappa, Z}$ corresponds to an eigenstate of the Dirac equation with nuclear charge $Z$. Eq.\ (\ref{unrenormalized_vapol}) contains infinite urenormalized charge which can be treated by computing the one-potential
\begin{align}
    U_1(r) = &\frac{4}{\pi}\int_{0}^{\infty}\dd k~j_0(kr_1) \sum_{\kappa}(2j_{\kappa}+1) \notag \\
             &\times\sum_{p}^+\sum_{q}^-\frac{\bra{p\kappa,0}j_0(kr)\ket{q\kappa,0}\bra{q\kappa,0}V(r)\ket{p\kappa,0}}{E_{p\kappa,0}-E_{q\kappa,0}},
    \label{onep}
\end{align}
where the free particle eigenstates $\ket{n\kappa,0}$ are obtained by solving the Dirac equation for $Z=0$. The charge divergency is then eliminated by subtracting Eq.\ (\ref{onep}) from Eq.\ (\ref{unrenormalized_vapol}); this is done term-by-term in $\kappa$ until the differences converge to zero. One can then obtain the energy shift $\Delta E_{n\kappa}$ of a reference state by taking the expectation value
\begin{align}
    \Delta E_{n\kappa} = \sum_{\kappa'} \int \dd \mathbf{x}~\psi_{n\kappa}^{\dagger}(\mathbf{x})\left[U_\text{VP}^{\kappa'}(r)-U_1^{\kappa'}(r)\right] \psi_{n\kappa}(\mathbf{x}).
\end{align}
The above expression corresponds to the vacuum polarization correction for all orders of $(Z\alpha)^{n\geq3}$. From here on, we will refer to this as the Wichmann-Kroll correction; however, it is also known as the many-potential vacuum polarization \cite{many_pot}.

\section{FINITE-BASIS SETS}
While it is possible to obtain solutions to the Dirac-Coulomb Hamiltonian by numerically solving the differential equation it leads to \cite{fin_element,fin_spur}, it is sometimes more convenient to work within the finite-basis approximation. Here, the radial functions are expanded in a set of basis functions that conform to the boundary conditions of the exact solution
\begin{align}
    \begin{bNiceMatrix}[cell-space-limits = 3pt]
     P_{n\kappa}(r) \\ Q_{n\kappa}(r)
    \end{bNiceMatrix}
    = \sum_{i}^N c_{\kappa;ni}
    \begin{bNiceMatrix}[cell-space-limits = 3pt]
        \varphi_{\kappa,i}^+(r) \\ \varphi_{\kappa,i}^-(r)
    \end{bNiceMatrix},
\end{align}
where $\varphi_{\kappa,i}^+(r)$ and $\varphi_{\kappa,i}^-(r)$ are the large- and small-component basis functions respectively. This transforms the problem into a matrix eigenvalue equation involving integrals over basis functions. The functions themselves are chosen such that:
\begin{itemize}
    \item they do not generate spurious energy solutions (variational  collapse) \cite{var_collapse}, and
    \item they respect the energy coupling between the large- and small-component radial solutions.
\end{itemize}
We will not discuss variational collapse here; however, a detailed analysis of it can be found in the work of Kutzelnigg \cite{var_kutz}.

While the second condition isn't necessary if one is only interested in positive (or negative) energy solutions, it is necessary for a complete representation of the energy spectrum, which includes energies above and below $mc^2$. A symmetric treatment of the large- and small-component radial functions that ensures the correct coupling is given by the dual kinetic balance (DKB) basis, first proposed by Shabaev \textit{et al} \cite{shabaev_DKB}. This basis expands the radial solutions as
\begin{align}
    \begin{bNiceMatrix}[cell-space-limits = 3pt]
     P_{n\kappa}(r) \\ Q_{n\kappa}(r)
    \end{bNiceMatrix} = \sum_i^{N^+} & c_{\kappa;ni}^+
    \begin{bNiceMatrix}[cell-space-limits = 3pt]
        \varphi_{\kappa,i}^+(r) \\ 
        \frac{1}{2c}\left(\frac{\dd}{\dd r}+\frac{\kappa}{r}\right)\varphi_{\kappa,i}^+(r)
    \end{bNiceMatrix} \notag \\ & + \sum_{i}^{N^-} c_{\kappa,ni}^-
    \begin{bNiceMatrix}[cell-space-limits = 3pt]
        \frac{1}{2c}\left(\frac{\dd}{\dd r} - \frac{\kappa}{r} \right) \varphi_{\kappa,i}^-(r) \\
        \varphi_{\kappa,i}^-(r)
    \end{bNiceMatrix}.
    \label{DKB}
\end{align}
For calculations of bound-state QED corrections, a further condition may be imposed on the basis set: charge-conjugation symmetry ($\mathcal{C}$-symmetry). A basis set is said to satisfy $\mathcal{C}$-symmetry if the charge-conjugation of an element returns another element of the set
\begin{align}
    \mathcal{C}\varphi_k\in \{\varphi_i\},~~~~\forall\varphi_k\in\{\varphi_i\}.
\end{align}
This is enforced in the DKB basis by ensuring that
\begin{align}
    \varphi^{\pm}_{\kappa,i}=\varphi^{\mp}_{-\kappa,i}.
\end{align}

Ivanov \textit{et al}.\ \cite{Ivanonv2024} have demonstrated the use of the DKB basis on hydrogenic systems. Their calculations of the Wichmann-Kroll correction show good agreement with the Green's integer method when a Gaussian basis is used. In this work, we utilize a DKB-like bases known as the charge-conjugated, kinetically-matched, Gaussian (CKG) basis.

\subsection{CKG-spinors}
$\mathcal{C}$-symmetry is built into the CKG-spinors, which were introduced by Grant and Quiney \cite{ckg_paper}. Here, the radial solutions are expanded as
\begin{widetext}
\begin{align}
    \begin{bNiceMatrix}[cell-space-limits = 3pt]
     P_{n\kappa}(r) \\ Q_{n\kappa}(r)
    \end{bNiceMatrix} = \sum_i^{n^+_\kappa} c_{\kappa;ni}^+ N^{+}_{\kappa,i}
    \begin{bNiceMatrix}[cell-space-limits = 3pt]
        \varphi_{\kappa,i}^+(r) \\ 
        \frac{1}{c+E_{+\kappa,i}/c}\left(\frac{\dd}{\dd r}+\frac{\kappa}{r}\right)\varphi_{\kappa,i}^+(r)
    \end{bNiceMatrix} + \sum_i^{n^-_\kappa} c_{\kappa;ni}^- N^{-}_{\kappa,i}
    \begin{bNiceMatrix}[cell-space-limits = 3pt]
        \frac{1}{c+E_{-\kappa,i}/c}\left(\frac{\dd}{\dd r} - \frac{\kappa}{r} \right) \varphi_{\kappa,i}^-(r) \\
        \varphi_{\kappa,i}^-(r)
    \end{bNiceMatrix},
\end{align}
\end{widetext}
where the parameter $E_{\pm\kappa,i}$ is chosen to be the magnitude of the expectation value of the free-particle kinetic energy, given by
\begin{align}
    E_{\pm\abs{\kappa},i} = c\sqrt{\langle p^2_{\pm\abs{\kappa},i}\rangle + c^2},
\end{align}
with 
\begin{align}
    \langle p^2_{+\abs{\kappa},i}\rangle = (2\abs{\kappa}+3)\lambda_{\abs{\kappa},i}, ~~
    \langle p^2_{-\abs{\kappa},i}\rangle = (2\abs{\kappa}+1)\lambda_{\abs{\kappa},i},
\end{align}
for basis exponent $\lambda_{\abs{\kappa},i}$. In the DKB basis of Eq.\ (\ref{DKB}) the energy parameter is set to $c^2$; however, given that calculations of QED corrections involve high-energy states, the free-particle energy is preferred. 

Grant \& Quiney \cite{ckg_paper} have demonstrated the convergence of energy eigenvalues using the CKG-spinors for hydrogenic systems. They have also demonstrated the convergence of the Wichmann-Kroll correction for one-electron point-nuclear systems.

\section{COMPUTATIONAL IMPLEMENTATION}
All calculations in the following section are performed using atomic units, with the speed of light set to ${c=137.03599911}$. All numerics are carried out in \texttt{FORTRAN90}, with the integration over the Wichmann-Kroll potential performed using the Gaussian quadrature routine in \texttt{QUADPACK}. For the nuclear potential, we utilize a Gaussian charge density of the form \cite{Dyall_book}
\begin{align}
    \rho^{\text{nuc}}(r) = Z \rho_0 \exp\left(-\zeta~r^2\right),
\end{align}
leading to a potential that is simply
\begin{align}
    V^{\text{nuc}} = -\frac{Z}{r} \erf\left(\sqrt{\zeta}~r \right),
\end{align}
with
\begin{align}
    \zeta = \frac{3}{2R},
\end{align}
where $R$ is the root mean square (rms) charge radius.

To obtain the mean-field electron density, we diagonalize the matrix
\begin{align}
    F_{\kappa} = H_{\kappa} + G_{\kappa},
    \label{fmat}
\end{align}
where $F_{\kappa}$ and $H_{\kappa}$ are the matrix representations of the Fock operator and Dirac Hamiltonian in the CKG-basis respectively, and
\begin{align}
    G_{\kappa} = 
    \begin{bNiceMatrix}[cell-space-limits = 3pt]
        G_{\kappa}^{++} & G_{\kappa}^{+-} \\
        G_{\kappa}^{-+} & G_{\kappa}^{--}
    \end{bNiceMatrix}.
\end{align}
Each submatrix of the $G$-matrix may be expanded as
\begin{align}
    G_{\kappa}^{\alpha\beta} = \sum_{\kappa',\gamma\lambda} G_{\kappa\kappa'}^{\alpha\beta\gamma\lambda}D_{\kappa'}^{\gamma\lambda},
\end{align}
where $D_{\kappa'}^{\gamma\lambda}$ is the density matrix. The construction of $G_{\kappa\kappa'}^{\alpha\beta\gamma\lambda}$ involves evaluating the matrix elements of the direct and exchange matrices, which requires solving two-electron integrals over CKG basis functions. The benefit of using a Gaussian type basis is that these integrals can be solved analytically \cite{kim_closed}. Further details on the DHF method for atoms, including the specific form of the two-electron integrals can be found in Refs.\ \cite{grant_book, grant_atm_str}.

It has been noted by Refs.\ \cite{Ivanonv2024, scooped} that maintaining charge-conjugation symmetry for vacuum polarization calculations requires numerical precision higher than 64 bits for floating point numbers. In particular, we have found that lowering $\beta$ below a value of approximately 1.35 results in a high degree of numerical error; therefore, we work in quadruple precision (128 bit floating point number) where possible.

For construction of the Fock matrix, we use the parameters:
\begin{align}
    \alpha=0.01,\quad\beta=1.35,\quad N=120,
\end{align}
in quadruple precision for the one-electron part of Eq.\ (\ref{fmat}). There is no requirement that the calculation of the mean-field density be carried out with the same parameters as the one-electron part; therefore, the self-consistent calculation of the mean-field utilizes the parameters 
\begin{align}
    \alpha=0.01,\quad\beta=1.50,\quad N=50,
\end{align}
which is sufficient for an accurate representation of the 1s$_{1/2}$ electron density. Lastly, we calculate the $G$-matrix in double precision initially and then promote it to quadruple precision to match the numerical type of the Dirac Hamiltonian. 

\section{RESULTS}
We now present the results of the hydrogen- and lithium-like energy corrections for systems of varying $Z$. For consistency, we compare both the one-electron and three-electron results with Sapirstein and Cheng \cite{Sapirstein_lilike_2003}. All calculations presented in this section are performed to a maximum partial wave of $\abs{\kappa}=10$.

Displayed in Table \ref{tab:wk_hlike} is the Wichmann-Kroll correction for the 1s$_{1/2}$ and $2s_{1/2}$ states of a few hydrogen-like systems. We find that the basis set results are consistently higher than Sapirstein and Cheng's for all systems. This is unsurprising considering the basis parameters chosen \cite{Ivanonv2024}. We note that the one-electron results can be improved by extrapolating $\beta$ to the complete basis set limit, a procedure first outlined by Schmidt and Ruedenberg \textit{et al.}\ \cite{completeness} and applied to the vacuum polarization by Benazzouk \textit{et al.}\ \cite{scooped}. However, for the purposes of comparing DHF results and screening, we have opted to make direct comparisons between both one-electron and three-electron systems for the same basis parameters. 
\begin{table}[ht]
\caption{\label{tab:wk_hlike}
Vacuum polarization correction for the 1s$_{1/2}$ and 2s$_{1/2}$ states of hydrogen-like systems. Results are compared to Saperstein and Cheng \cite{Sapirstein_lilike_2003}, who use a Fermi nucleus for the potential.}
\begin{ruledtabular}
\begin{tabular}{cccccc}
& & \multicolumn{2}{c}{This work} & \multicolumn{2}{c}{\cite{Sapirstein_lilike_2003}}\\
\cline{3-4}\cline{5-6}
\textrm{$Z$}&\textrm{$R$ (fm)}& 1s$_{1/2}$& 2s$_{1/2}$ & 1s$_{1/2}$ & 2s$_{1/2}$ \\
\colrule
    70 & 5.229 & 0.03091 & 0.00448 & 0.03029 & 0.00438 \\
    80 & 5.464 & 0.07244 & 0.01106 & 0.07144 & 0.01089 \\
    90 & 5.707 & 0.15922 & 0.02590 & 0.15744 & 0.02556 \\
    92 & 5.751 & 0.18476 & 0.03058 & 0.18254 & 0.03013 \\
   100 & 5.886 & 0.33680 & 0.05902 & 0.33397 & 0.05845 \\
\end{tabular}
\end{ruledtabular}
\end{table}

In Table \ref{tab:wk_lilike} we present the Wichmann-Kroll correction for various lithium-like systems. For the electron screening, Sapirstein and Cheng utilize a Kohn-Sham potential \cite{atom_spectra} rather than a Hartree-Fock potential \cite{sap_ks}; this is likely the source of the discrepancy between vacuum polarization screening between the two methods. Nevertheless, we find that the calculation of the Wichmann-Kroll correction for lithium-like systems produced in this work agree with those of Sapirstein and Cheng to within 1\%. 
\begin{table}[ht]
\caption{\label{tab:wk_lilike}
Dirac-Hartree-Fock results for the vacuum polarization of the 2s$_{1/2}$ state of lithium-like systems. The electron screening is taken as the difference between the 2s$_{1/2}$ vacuum polarization of the lithium-like and hydrogen-like systems.}
\begin{ruledtabular}
\begin{tabular}{cccccc}
& & \multicolumn{2}{c}{This work} & \multicolumn{2}{c}{\cite{Sapirstein_lilike_2003}}\\
\cline{3-4}\cline{5-6}
\textrm{$Z$}&\textrm{$R$ (fm)}& 2s$_{1/2}$& Screening & 2s$_{1/2}$ & Screening \\
\colrule
    70 & 5.229 & 0.00423 & 0.00025 & 0.00420 & 0.00018 \\
    80 & 5.464 & 0.01051 & 0.00055 & 0.01044 & 0.00045 \\
    90 & 5.707 & 0.02468 & 0.00122 & 0.02455 & 0.00101 \\
    92 & 5.751 & 0.02915 & 0.00143 & 0.02900 & 0.00113 \\
   100 & 5.886 & 0.05637 & 0.00265 & 0.05627 & 0.00217 \\
\end{tabular}
\end{ruledtabular}
\end{table}

It would be instructive to observe how the three-electron energy correction varies with $\beta$; however, calculating the $G$-matrix in double precision places a lower limit on this value as numerical error becomes a problem at $\beta<1.35$. To produce a meaningful extrapolation Bennazouk \textit{et al.}\ (Fig.\ 15 Ref.\ \cite{scooped}) go down to $\beta=1.19$. This cannot be achieved without using quad precision for both the Hamiltonian and $G$-matrix, which places a significant overhead on computation time. 
\newline
\section{CONCLUSION}
In this work, we have calculated the Wichmann-Kroll correction for hydrogen-like and lithium-like systems using a Gaussian type basis that satisfies $\mathcal{C}$-symmetry by construction. We find good agreement between the finite-basis approach used in this work and the Green's function approach taken by Sapirstein and Cheng \cite{Sapirstein_lilike_2003}. However, a more general class of potentials can be analyzed using the finite-basis scheme without adding too much complexity.

It is clear that the limiting factor in the accuracy of these calculations is the computational linear dependence of the basis set, which places a lower bound on the value of numerical value of $\beta$. One can reduce this bound by increasing the numerical precision; however, this adds a large overhead to the execution time of the program, particularly for two-electron matrix elements. Nevertheless, the use of Gaussian basis sets opens the door to \textit{ab initio} QED calculations of molecular systems, as well as more complicated atomic systems.
%\newpage

\bibliography{apssamp}

\end{document}